\documentclass[aps,prl,twocolumn,groupedaddress]{revtex4-2}

\usepackage{graphicx}
\usepackage{amsmath,amsfonts,amssymb}
\usepackage{bm} 

\begin{document}

\title{Complex Memory Formation in Frictional Granular Media}
\author{D. Candela}
\email[]{candela@physics.umass.edu}
\affiliation{Physics Department, University of Massachusetts, Amherst MA 01003}

\date{\today}

\begin{abstract}
	Using numerical simulations it is shown that a jammed, random pack of soft frictional grains can store an arbitrary waveform that is applied as a small time-dependent shear while the system is slowly compressed.
	When the system is decompressed at a later time, an approximation of the input waveform is recalled in time-reversed order as shear stresses on the system boundaries.
	This effect depends on friction between the grains, and is independent of some aspects of the friction  model.
	This type of memory could potentially be observable in other types of random media that form internal contacts when compressed.
\end{abstract}

\maketitle

	There are many forms of memory in condensed matter, i.e. ways in which inputs applied over one time period can appear as measurable outputs at later times~\cite{keim19}.
	Here it is shown using simulations  that the elastic properties of a granular medium -- a random pack of $\mu$m or larger size particles with contact interactions -- can be used to encode arbitrary waveform data, store it for long times, and  (imperfectly) recall the data when desired.
	Similar waveform memory might be observable in other systems that share key properties with granular media such as fiber nests~\cite{bhosale22}, fiber bundles and yarns~\cite{panaitescu18,seguin22}, textiles~\cite{poincloux18}, and crumpled sheets~\cite{matan02,cambou11,lahini17}.
	This type of effect is broadly interesting as it shows that a detailed record of past events can,  surprisingly, sometimes be stored and read out from ``ordinary'' random media not designed for the purpose.

	A variety of memory effects have been found previously in amorphous many-particle systems.
	For example, multiple shear values can be stored via cyclic training in both suspensions~\cite{paulsen14} and model glasses~\cite{fiocco14,mukherji19,keim20}.
	Granular media can also be trained by cyclic shearing~\cite{toiya04,mueggenburg05,ren13,zhao22} and store information on shear history in the fabric of force chains~\cite{zhao19}.
	Nonlinear acoustic modes of granular media give rise to time-domain echoes~\cite{burton16}.

	The memory effect reported here differs from most of these earlier results as the input data are stored and later recalled with a single cycle of the control variable.
	The memories are complex in the sense that they approximately store an entire time-dependent waveform that could eventually represent,  for example, a spoken word.
	In simulations the memories are stored indefinitely, but in physical systems creep~\cite{dijksman22} might limit storage times.

	Many of the unusual mechanical and acoustic properties of granular media  can be traced to the network of contacts between grains~\cite{ohern03,somfai05,liu10}.
	Additional contacts are formed as the sample is compressed beyond the jamming point, and a large system always includes contacts close to forming or separating~\cite{makse00,agnolin07b}.
	When there is contact friction the forces depend upon the path by which the granular system reaches a given state, creating possibilities for memory effects~\cite{mindlin53,elata96,johnson97}.

 \begin{figure}
 \includegraphics[width=\linewidth]{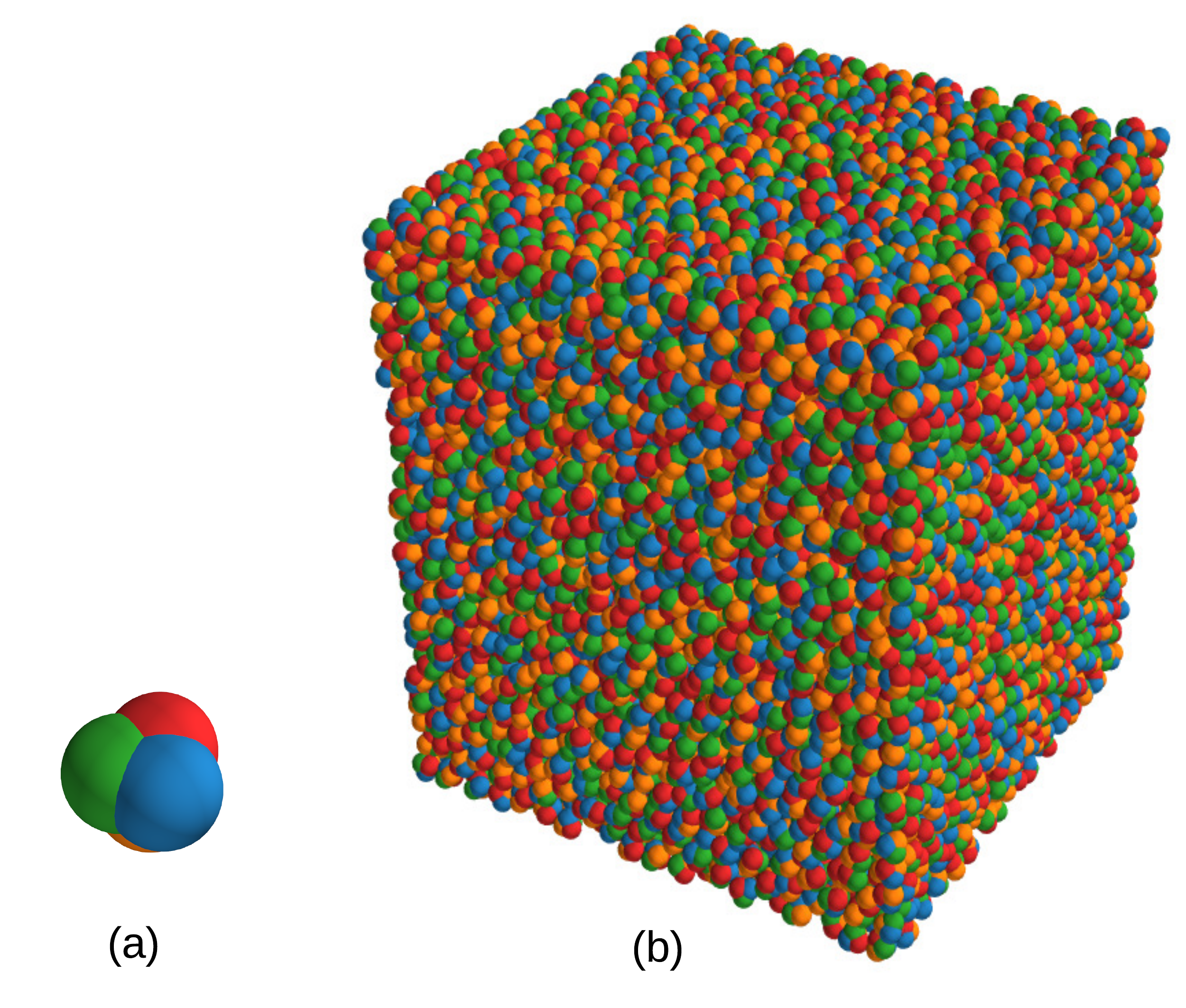}
 \caption{\label{fig1}(a) A single grain, which has the exterior shape of four partially overlapping spheres.
 	The sphere centers are at the vertices of a tetrahedron with edge length 40\% of the sphere diameter.
 	(b) Random pack of $10\,240$ such grains (confining walls not shown).}
 \end{figure}

	\emph{Simulated system.---}Parameters were chosen that could be recreated in physical experiments.
	A non-spherical grain shape was used, modeled for simplicity~\cite{gallas93,nguyen19} by four tetrahedrally-arranged partially overlapping  spheres of radius $R=0.5$~mm (Fig.~\ref{fig1}(a)) with the properties of silicone rubber (density $\rho = 1.2\times10^3$~kg/m$^3$, Young modulus $E=1.0\times10^7$~Pa, Poisson ratio $\nu=0.49$, friction coefficient $\mu \approx 1.0$).
	A soft elastomer like this can withstand strains of several percent, as used in the simulations.
	A characteristic time from these parameters is $t_c = R(\rho/E)^{1/2}$, which is about one-tenth the period of the highest-frequency vibrational mode of a moderately compressed pack of these grains.

	As detailed below the contact forces are modeled by a Hertzian repulsive normal force along with three successively more realistic friction force models denoted H, M1, M2.
	Apart from M2 these models have commonly been used to simulate granular media.

	Conventional discrete element method (DEM) methods~\cite{cundall79,silbert01,poschel05,luding08, matuttis14} were used to integrate the equations of motion for $n_g$ grains with these forces, confined by frictionless walls.
	The data shown here are for $n_g=10\,240$, but the memory effect is visible at reduced fidelity for $n_g$ as small as 325.
	An integration time step $\Delta t \leq  (0.4)t_c$ gave numerical stability, requiring of order $10^6$ steps.

 \begin{figure}
 \includegraphics[width=\linewidth]{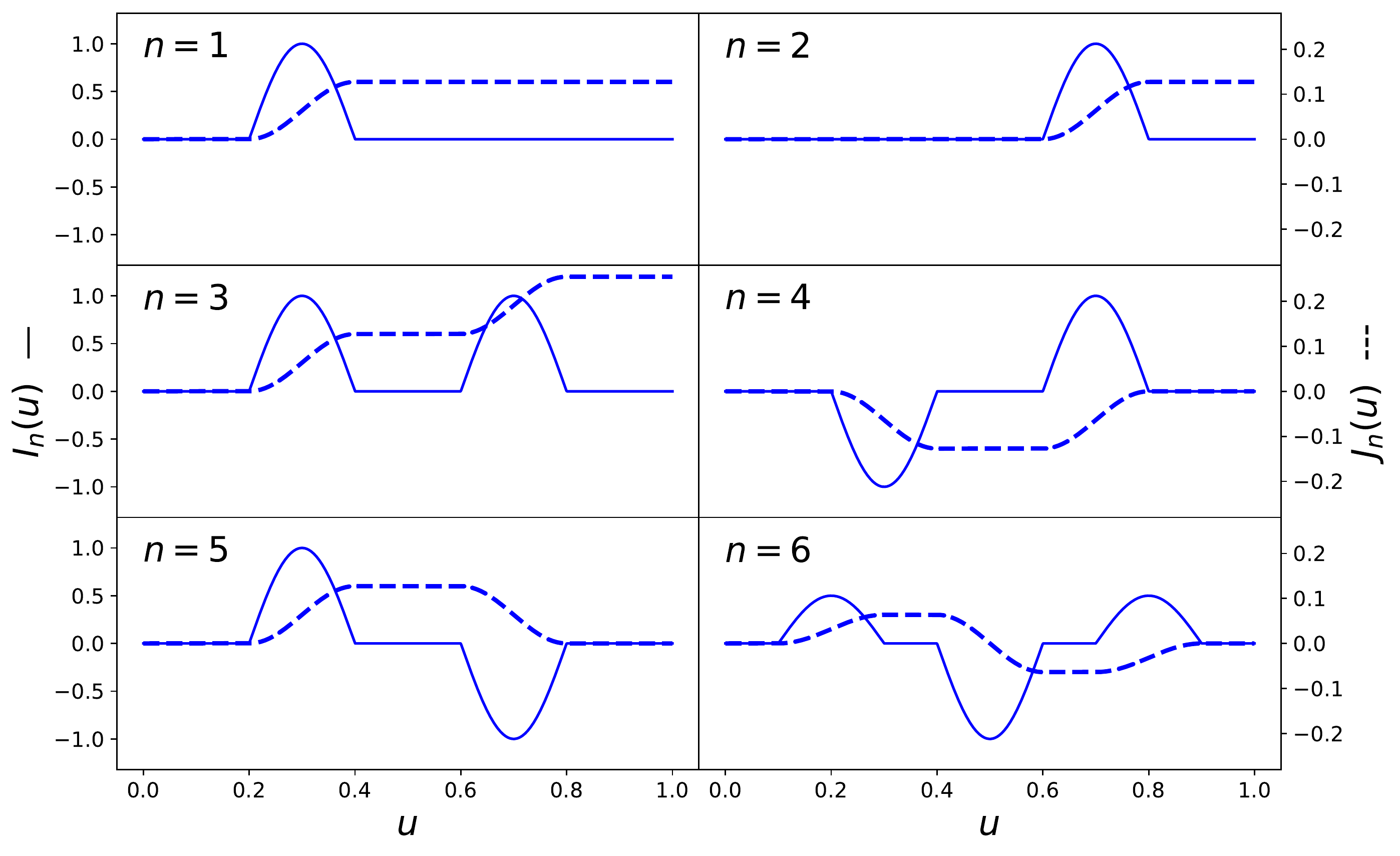}
 \caption{\label{fig2} Six  input waveforms $I_n(u)$ used for memory trials (additionally $I_0(u)=0$ was used).
 	Each waveform is composed of up to three half-cycle cosine curves.
 	 Shear strain proportional to $I_n(u)$ was applied to the sample (solid curves), but the shear stress recalled later was found to be proportional to the integral of the strain $J_n(u) = \int_0^u I_n(v)dv$ (dashed curves).
 }
 \end{figure}

 \begin{figure}
 \includegraphics[width=\linewidth]{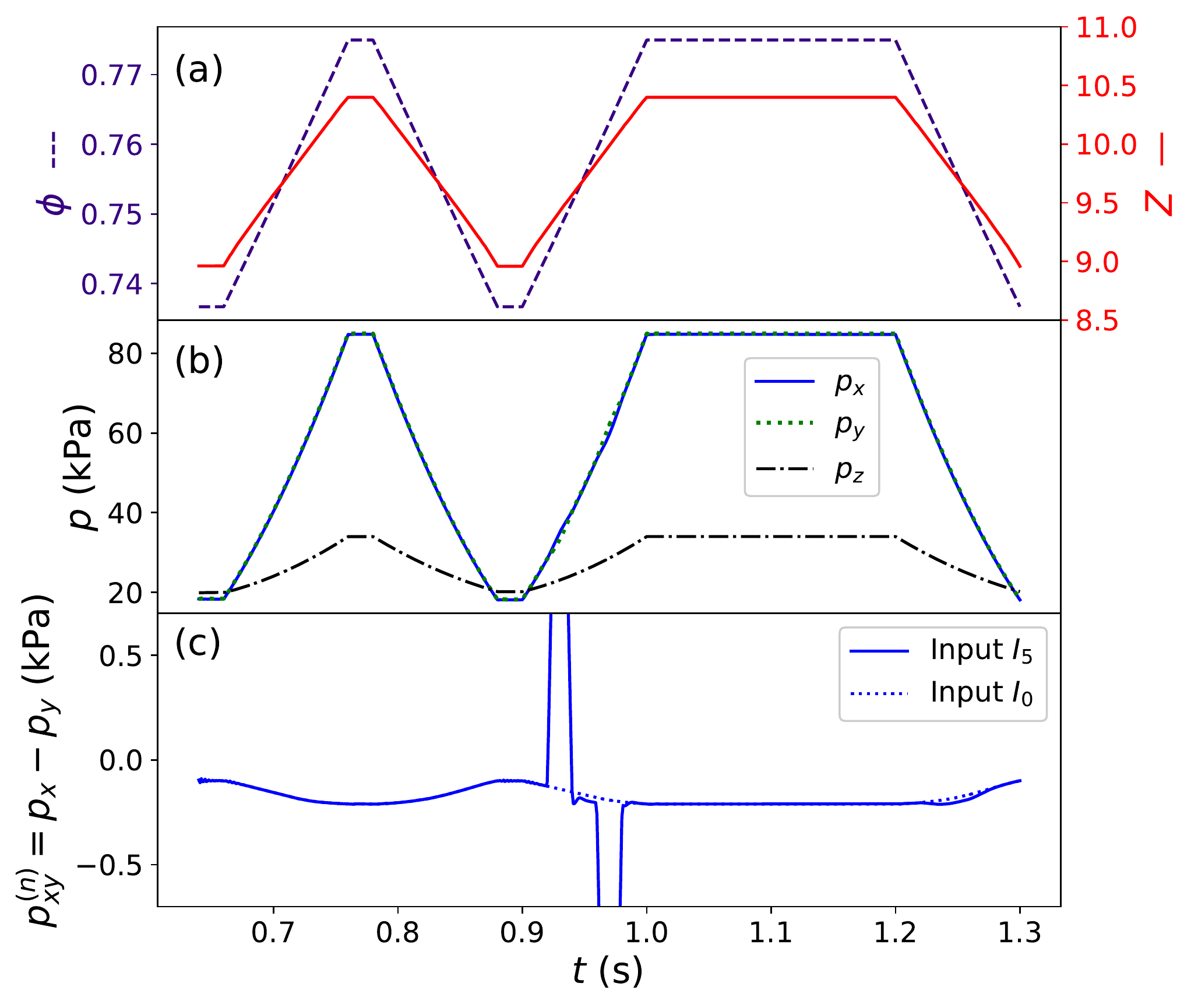}
 \caption{\label{fig3} Quantities versus time $t$ for a memory experiment simulation using input signal $I_5$.
 	The the sample preparation period with zero friction $t<0.64$\,s is not shown; friction is turned on for the entire period shown here.
	After a preparatory compression cycle (0.68\,s - 0.88\,s), the sample is compressed while the input signal $I_5$ is applied as an $x$-$y$ shear strain (0.9\,s - 1.0\,s), held compressed for a storage period (1.0\,s -1.2\,s), and finally decompressed to read out the memory response (1.2\,s - 1.3\,s).
 	(a)~Sample filling factor $\phi(t)$ and average coordination number $Z(t)$.
 	Note tetrahedral particles typically pack more densely than spheres~\cite{hajiakbari09}.
 	(b)~Pressures on the $x$, $y$, and $z$ walls.
 	The responses of $p_x,p_y$ to the input signal are barely visible during the encoding period 0.9\,s -1.0\,s.
 	(c)~Difference between $p_x$ and $p_y$ on an expanded vertical scale, shown both when input $I_5$ is applied and when the zero-shear input $I_0$ is applied.
 	Now the response to $I_5$ during the encoding period is readily visible, while differences between the responses to $I_5$ and $I_0$ are barely visible during the readout period 1.2\,s - 1.3\,s .}
 \end{figure}

	\emph{Sample preparation.---}An initial simulation was used to prepare the packed granular sample.
	The grains were placed in a rectangular box on a low-density lattice with random velocities, and allowed to evolve to randomize positions and orientations~\cite{agnolin07a}.
	Then with the $+z$ wall  free to move an external pressure $p_z = (2\times10^{-3})E$ was used compress the grains into an approximately cubical pack, Fig.~\ref{fig1}(b).
	After the system to came nearly to rest the $\pm z$ walls were fixed in place while the $\pm x, \pm y$ walls were used to compress and decompress the sample two times mimicking the compression cycles used later in the experiment simulations.
	 The grain-grain friction coefficient $\mu$  was set to zero during this entire sample preparation procedure~\cite{makse99,thornton00,agnolin07a}.

	 \emph{Memory experiments.---}A single sample prepared as above was used as the starting point for multiple  memory-experiment simulations using different shear input waveforms.
	 For these simulations $\mu$ was set to the chosen value (1.0 except as noted).
	To exhibit the memory effect, the grain pack was slowly and linearly compressed, while simultaneously applying an arbitrary input waveform as a small shear strain.
		The compression was applied by moving the four walls $\pm x$, $\pm y$ inward simultaneously and linearly in time over a period $t_0 = (1.8\times 10^4)t_c$, to  change the sample volume by $\Delta V/V = -\delta_0 = -0.05$.
	The average coordination number (contacts on a grain) $Z$ increased from 9.0 to 10.4 as the sample was compressed, remaining between the frictional ($Z=4$) and frictionless ($Z=12$) isostatic values~\cite{donev07,henkes10,isostatic}.
	The the inertial number $I = \dot{\epsilon}\sqrt{m/Dp}$ (with $\dot{\epsilon}$ the strain rate, $m,D$ the grain mass and diameter, and $p$ the pressure) was always less  than $4\times10^{-5}$, giving a nearly quasistatic compression~\cite{dacruz05,agnolin07b}.

	The compression was parameterized by a variable $u(t)$ that went from zero to one as the sample was compressed, then back to zero when the sample was decompressed.	
	The sample was compressed and decompressed once without applied shear.
	Then, during the course of a final compression, an input was applied by small additional movements of the $\pm x$ walls inward while moving the $\pm y$ walls outward (or vice versa), so as to create a pure shear of the sample $\gamma(u) = \gamma_0 I_n(u)$.
	Here $\gamma_0 = 10^{-3}$ set the scale of the shear strain and  $I_n(u), n=0\dots 6$ were seven different input waveforms with $-1\leq I_n(u) \leq 1$ (Fig.~\ref{fig2}), used for seven separate experiment simulations.
	The $I_n(u)$ were chosen as simple waveforms to  minimally test the independent recall of multiple inputs in a single experiment.
	
	It was found that the recalled shear-stress signal when the sample was decompressed was proportional not directly to the applied shear $I_n(u)$ but rather to its integral $J_n(u) = \int_0^u I_n(v)dv$.
	This emerges as well from a rough explanation for the memory effect presented below.

 \begin{figure}
 \includegraphics[width=\linewidth]{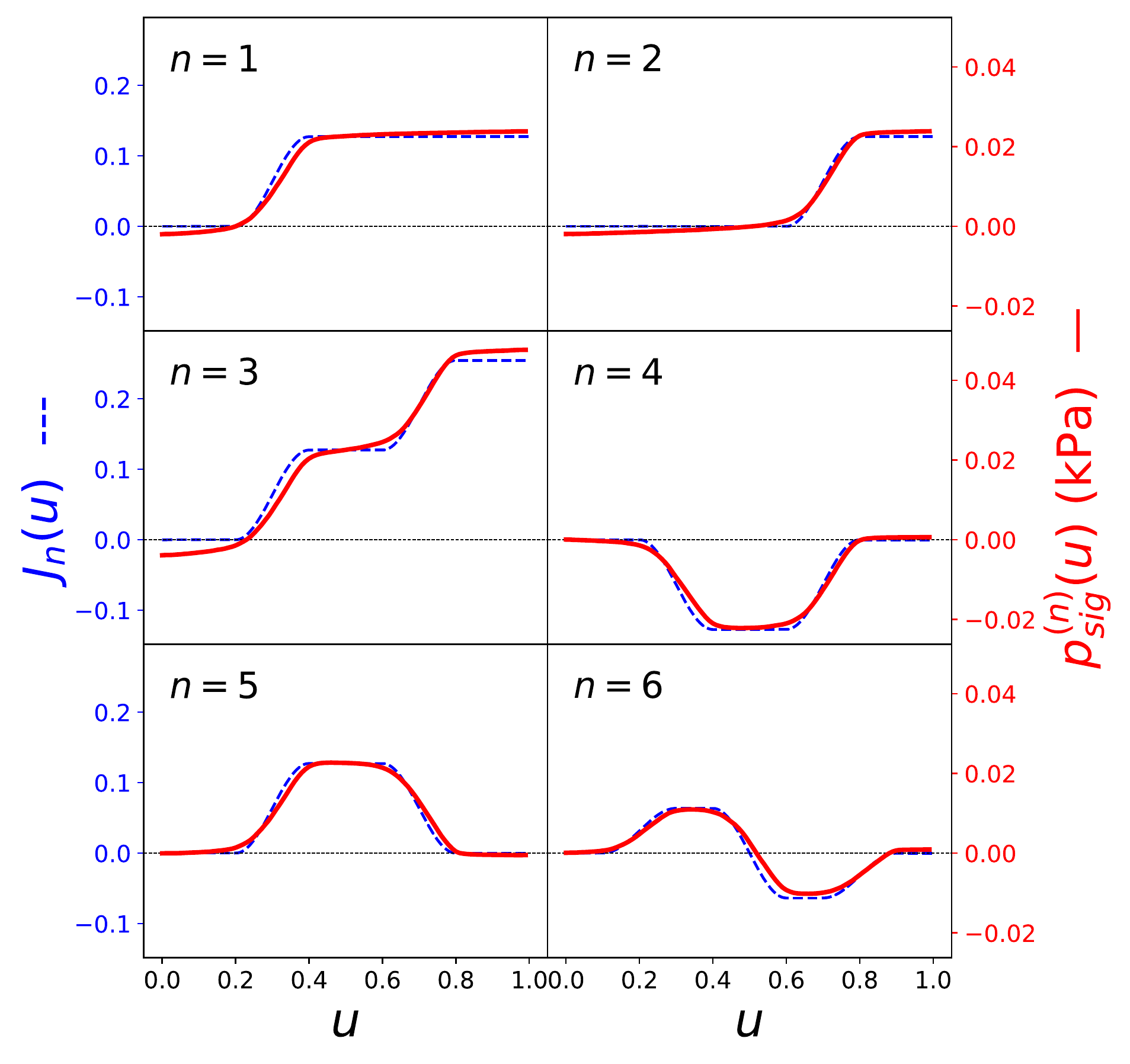}
 \caption{\label{fig4} Recalled signal $p_\mathit{sig}^{(n)}(u) = -(p_{xy}^{(n)} - p_{xy}^{(0)})$ when the sample is decompressed, for the six input signals $n=1\dots6$ shown in Fig.~\ref{fig2} (solid lines).
 	For comparison the dashed lines show the corresponding  integrated shear inputs $J_n(u)$ applied earlier when the sample was compressed.
 	A single gain factor $\mathcal{G} = 1.77\times10^5$\,Pa was computed to minimize the least-squares difference between $\mathcal{G}\gamma_0 J_n(u)$ and $p_\mathit{sig}^{(n)}(u)$ summed over all six signals, and used to scale all plots equally. }
 \end{figure}
 
	\emph{Simulation results.---}Figures~\ref{fig3}(b,c) show the pressures $p_x$, $p_y$, $p_z$ measured on the sample walls during a typical simulation run, along with the shear stress $p_{xy}^{(n)} = p_x - p_y$ which has the symmetry of the applied input signals.
	While the response of $p_{xy}^{(n)}$ to the input signal $I_n(u)$ during compression is clearly visible, the recalled response when the sample is later decompressed is barely visible Fig.~\ref{fig3}(c) due to the background signal observed even when no input signal is applied.
	This background reflects the $x$-$y$ asymmetry of a specific random pack.
	
	In Fig.~\ref{fig4} the zero-input background (measured in a separate simulation run) is subtracted to give the processed response signal $p_\textit{sig}^{(n)} = -(p_{xy}^{(n)} - p_{xy}^{(0)})$.
	It can be seen that the response $p_\textit{sig}^{(n)}(u)$ is nearly proportional to the input signal $J_n(u)$, demonstrating the postulated granular memory effect.
	A single parameter $\mathcal{G}$ was adjusted to minimize the integrated least-squares difference between $\mathcal{G}\gamma_0J_n(u)$ and $p_\textit{sig}^{(n)}(u)$ summed over the entire suite shown in Fig.~\ref{fig4}~\cite{supplementary}.

	Discrepancies between the recalled and input signals are also visible in Fig.~\ref{fig4}.
	These suggest limits on the complexity of the memory that can be stored, possibly related to the size of the granular system and the amplitude and/or frequency composition of the input signals.
	This will be the subject of future investigations.
	
	During the storage period between encoding and recalling the signals the maximum grain velocity decayed exponentially towards the numerical noise floor, implying the signals could be stored indefinitely.
	
\begin{figure}
 \includegraphics[width=\linewidth]{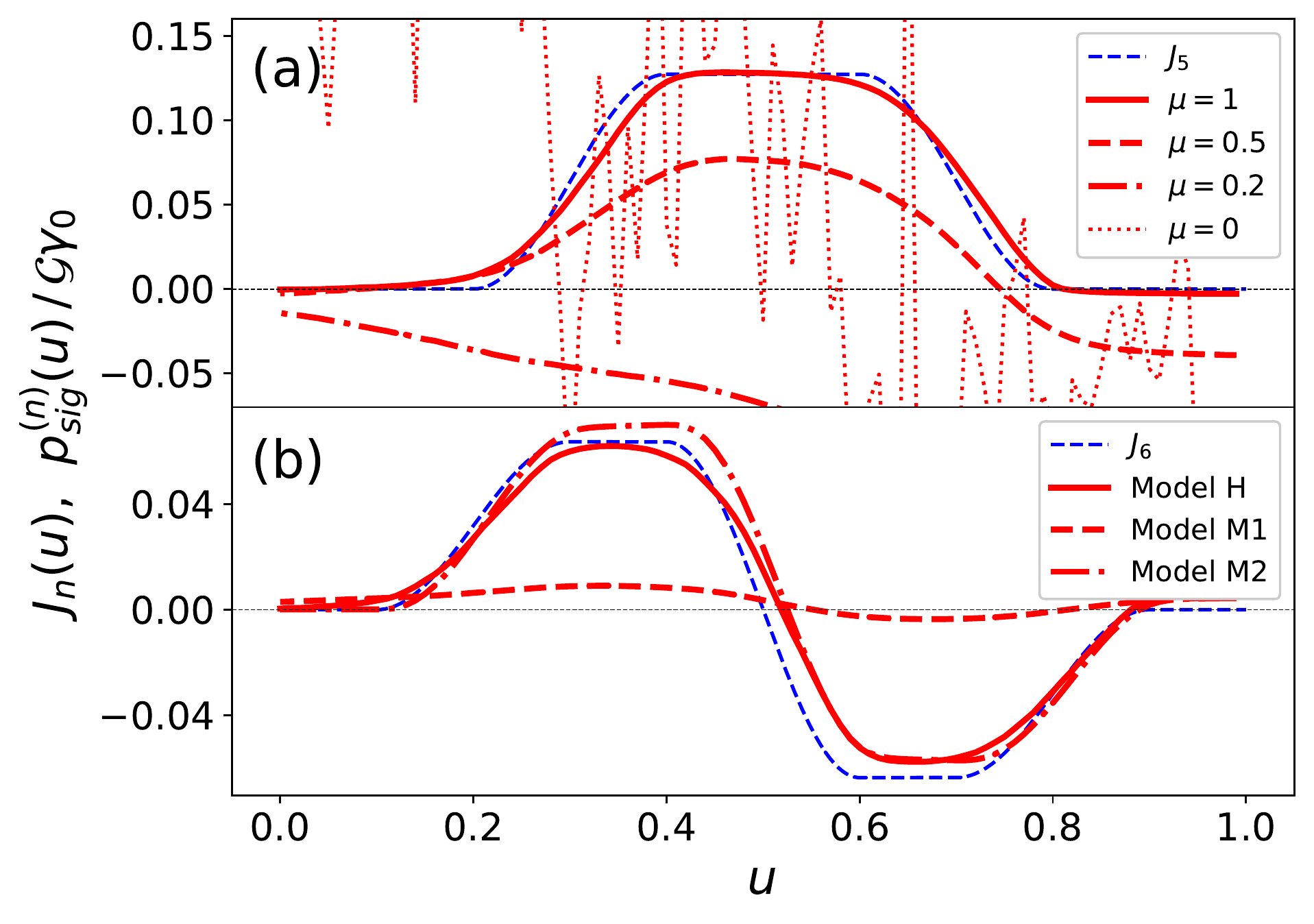}
 \caption{\label{fig5} Effects of varying the friction coefficient and the friction force model.
 	In each case the response $p_\mathit{sig}^{(n)}(u)$ to to a single input signal  ($I_5$ or $I_6$) is shown, but the gain $\mathcal{G}$ was adjusted as in Fig.~\ref{fig4} to minimize the errors summed over all six input signals.
 	Details of this fitting procedure and the resulting $\mathcal{G}$ values are in Supplementary Material~\cite{supplementary}.
 	(a)~Effect of varying the grain-grain friction coefficient $\mu$.
 	The memory effect is essentially absent for $\mu\leq0.2$, and when $\mu=0$ numerous particle rearrangements make $p_\mathit{sig}^{(n)}(u)$ noisy.
 	(b)~Effect of varying the transverse force model.
 	The simplest model H and the most realistic model for viscoelastic spheres M2 show similar memory effects, while the intermediate model M1 shows very little memory.}
 \end{figure}

	\emph{Dependence on the friction coefficient.---}
	Additional simulations to be reported elsewhere were carried out with frictional sample preparation and/or spherical grains.
	It was found that both frictionless sample preparation and non-spherical grains as used for the results reported here tended to make the memory effect more visible, by reducing the frequency of large grain movements during compressions~\cite{murphy19}.
	Similarly it was found in Ref.~\cite{kramar21} that large grain movements destroy memory when a granular medium is sheared.
	However the parameter that appears most directly to control the ability to store memories is the friction coefficient $\mu$, Fig.~\ref{fig5}(a).
	This suggests that friction at contacts is the ultimate physical origin of this phenomenon.

	\emph{Heuristic explanation of memory effect.---}A rough explanation is as follows:
	At the point $u$ during the compression at which a particular contact is formed, the grains coming into contact are displaced relative to one another by an amount proportional to the shear strain $\gamma_0 I_n(u)$ applied to the sample, and when this input strain is later removed there will be a corresponding transverse stress in the contact due to friction.
	At the end of the compression, the externally measurable wall stress $p_{xy}^{(n)}(u=1)$ should have contributions proportional to all such contact stresses, i.e. to $ \gamma_0\int_0^1 I_n(u) du = \gamma_0 J_n(1)$.
	During the decompression each contact stress is relieved at approximately the same point $u$ at which the contact was formed, giving a contribution to $p_{xy}^{(n)}(u) \propto -\gamma_0 J_n(u)$.
	
	For a quantitative theory, it would be necessary to connect the macroscopic applied strain $\gamma_0 I_n(u)$ to the distribution of transverse grain movements at contacts, and similarly to connect grain-scale friction forces to the macroscopic wall stress $p_{xy}^{(n)}(u)$.
	This is nontrivial due to non-affine grain motion and the creation of contacts as the sample is compressed \cite{makse99}.
	
	\emph{Dependence on the friction model.---}To check if the memory effect could be an artifact of the friction model, simulations were carried out using three different models.	
	A Hertzian repulsive normal force~\cite{brilliantov96,schwager08} was used for all three models,
\begin{equation}\label{fn}
	f_n = \mbox{max}(0, k_n p^{3/2} + \gamma_n p^{1/2}\dot{p})
\end{equation}
as appropriate for viscoelastic rather than plastic grains \cite{walton93,vuquoc99,thornton13}.
	Here $p$ is the normal overlap of two grains, $\dot{p}$ its rate of change, $k_n=2^{-3/2}(4/3)R^{1/2}E/(1-\nu^2)$, and  $\gamma_n/k_n = (0.23)t_c$ was used.
	Rolling at contacts is important in granular compressions~\cite{kuhn04,benson22}; the rolling and twisting resistances were set to zero.

	Friction models H, M1, M2 all include transverse elastic and damping force vectors $\mathbf{f}_e$, $\mathbf{f}_d$ with the magnitude of the total transverse force limited by the Coulomb criterion $|\mathbf{f}_e + \mathbf{f}_d| < \mu f_n$ using the algorithm of
 Ref.~\cite{luding08}.
	
	Model H (``Hooke'') uses a linear spring and dashpot
\begin{equation}\label{fth}
	\mathbf{f}_e = k_H \bm{\sigma},\ \ \ \mathbf{f}_d = \gamma_H \dot{\bm{\sigma}}
\end{equation}
with  $\bm{\sigma}$ is the accumulated vector sliding motion between the two grain surfaces and $\dot{\bm{\sigma}}$ its rate of change.
	Without the damping term this is the original friction model of Ref.~\cite{cundall79}.
	Here $k_H$ was set to a typical inverse transverse compliance~\cite{johnson85} from model M1 below, and $\gamma_H$ was set using $\gamma_H/k_H = (0.3)\gamma_n/k_n$.
	The data shown in Figs.~\ref{fig3}-\ref{fig4} and \ref{fig5}(a) are for this simplest friction model H.

	Model M1 (``Mindlin-1'') improves upon model H by making the transverse compliance dependent upon the normal overlap $p$, using a linearized, no-slip version~\cite{johnson97} of the transverse force calculated by Mindlin and Deresiewicz for contacting elastic spheres \cite{mindlin49,mindlin53}.
	In this model the elastic force is accumulated using
\begin{equation}\label{ftm1}
	\Delta\mathbf{f}_e = k_Mp^{1/2} \Delta\bm{\sigma},\ \ \ \mathbf{f}_d = \gamma_M p^{1/2}\dot{\bm{\sigma}}
\end{equation}
	with $k_M = 3k_n(1-\nu)/(2-\nu)$ ~\cite{johnson85,makse99}.
	As with model H $\gamma_M/k_M = (0.3)\gamma_n/k_n$ was used.
	To allow for the loss of stored elastic energy when $p$ decreases, an approximation due to Walton is used \cite{agnolin07a,thornton13,lammpsmindlin}: the elastic force $\mathbf{f}_e$ is reduced proportionally to $p^{1/2}$.
	This model M1 has frequently been used for DEM simulations of frictional granular matter~\cite{makse99,makse04,agnolin07a,lammpsmindlin}.
	
	Interestingly, when the nominally more realistic model M1 is substituted for model H, the memory effect reported here nearly disappears, Fig.~\ref{fig5}(b).
	This can be traced to the approximation for the reduction of elastic energy, which effectively assumes that the current value of $\mathbf{f}_e$ results entirely from sliding motion at the current value of $p$ and thus erases memory of the degree of compression at which shear strains were applied to the sample.
	
	Model M2 (``Mindlin 2'') avoids this approximation by directly computing the change of $\mathbf{f}_e$ with decreasing overlap $p$ from the linearized Mindlin model of Ref.~\cite{johnson97}.
	One way to do this (used here) is to represent the transverse elastic force in a contact as an integral over contributions from sliding at different values of $q = p^{1/2}$, i.e. $\mathbf{f}_e = \int_0^\infty \mathbf{f}(q) dq$.
	The first of Eqs.~\ref{ftm1} is implemented by adding $k_M \Delta\bm{\sigma}$ to $\mathbf{f}(q)$ in the interval $0 < q < p^{1/2}$, and when $p$ decreases $\mathbf{f}(q)$ is set to zero for $q>p^{1/2}$.
	Note this is not  a new friction model, but rather a more accurate representation than M1 of the transverse force law of Ref.~\cite{johnson97}.
	Model M2 has not typically been used for DEM simulations because it requires keeping a (suitably discretized) vector-valued function $\mathbf{f}(q)$ of history information for each contact.
		This complex path history retention in each contact is an inherent property of the Mindlin-Deresiewicz theory~\cite{mindlin53,johnson97}, but as the Model H results show it is not required for the bulk memory effect reported here.

	With the more-faithful linearized Mindlin model M2, the memory effect is very similar to that seen for the simplest model H (Fig.~\ref{fig5}(b)).
	Although friction models other than the three considered here have been used \cite{walton86,shafer96,poschel05} models H and M2 are sufficiently different to suggest that the memory effect is insensitive to details of the model, provided it does not explicitly erase memories as M1 does.
	
	The three models H, M1, M2 essentially assume grain-scale elasticity coupled with more microscopic Coulomb-law friction at contacting surfaces~\cite{johnson85}.
	This should be valid for large elastomer grains, but harder, more sand-like granular media might not have this separation of scales.
	Recent work has explored microscopic, asperity-based models of friction for jammed granular media~\cite{papanikolaou13,ikeda20}, and it will be interesting to see if the memory effect reported here persists for such models.
	Also harder grain materials typically have smaller yield strains.
	Modeling them would require reducing the compression factor $\delta_0$, probably necessitating larger simulations to observe the shear strain memory as a bulk effect.
	
	\emph{Conclusions.---}It is found via simulations that random packs of soft, frictional grains effectively store for indefinite times input signals applied as small shear waveforms while the sample is slowly compressed, with the signals recalled when the sample is eventually decompressed.
	
	The proposed mechanism of memory formation is similar to holography, in that a local nonlinear phenomenon (contact friction) is used to record correlations between the input signal and a reference signal (the grain motions during compression).
	Re-application of the reference signal at a later time (by decompressing the sample) allows the memory to be read out.

	The property of granular media that appears essential to such memory formation---the creation of internal, frictional contacts upon compression---exists for many other types of random media ~\cite{bhosale22,panaitescu18,seguin22,poincloux18,matan02,cambou11,lahini17} which suggests that similar memory effects might be found more generally.
	
\begin{acknowledgments}
	This work was completed in part with resources provided by the University of Massachusetts' Green High Performance Computing Cluster (GHPCC).
\end{acknowledgments}

\bibliography{gmbib}

\begin{thebibliography}{60}
\bibitem [{iso()}]{isostatic}%
  \BibitemOpen
  \href@noop {} {}\bibinfo {note} {It is not clear how relevant the concept of
  isotaticity is to the simulations reported here, as the grains are
  nonspherical, nonconvex, and compressible.}\BibitemShut {Stop}%
\end{thebibliography}

\end{document}


\section*{Supplementary material for\\
``Complex memory formation in frictional granular media''}

\subsection*{Determination of the gain and fidelity of the memory effect}
	It was found in the simulations that the shear stress $p_\textit{sig}^{(n)}(u)$ measured while the sample was decompressed was approximately proportional to the integrated shear strain $\gamma_0 J_n(u)$ applied earlier while the sample was compressed,
\begin{equation}\label{pggj}
	p_\textit{sig}^{(n)}(u) \approx \mathcal{G} \left( \gamma_0 J_n(u) \right)
\end{equation}
for $n=1\dots N$, with  $N=6$ for the data shown in the main text.
	The proportionality factor or ``gain'' $\mathcal{G}$ has units of pressure, and on dimensional grounds must itself be proportional to the grain modulus $E$.	
	The gain $\mathcal{G}$ was determined by minimizing squared error in Eq.~\ref{pggj} integrated over $u$ and summed over input signals,
\begin{equation}\label{eesq}
	\mathcal{E}^2 = \sum_{n=1}^N \int_0^1 \left( p_\textit{sig}^{(n)}(u) - \mathcal{G} \gamma_0 J_n(u) \right)^2  du.
\end{equation}
	Considering a set of $N$ $u$-dependent functions to be a  vector and defining the dot product
\begin{equation}
	\{F_n(u)\}_{n=1\dots N} \leftrightarrow \mathbf{F},\ \ \ \mathbf{F}\cdot\mathbf{G} = 
	 \sum_{n=1}^N \int_0^1 F_n(u) G_n(u) \, du
\end{equation}
the total error squared is
\begin{equation}\label{eesq}
	\mathcal{E}^2 = (\mathbf{p}_\textit{sig} - \mathcal{G}\gamma_0\mathbf{J}) \cdot
	(\mathbf{p}_\textit{sig} - \mathcal{G}\gamma_0\mathbf{J}).
\end{equation}
	Minimizing $\mathcal{E}^2$ by setting $d\mathcal{E}^2/d\mathcal{G} = 0$ finally yields
\begin{equation}\label{gg}
	\mathcal{G} = \frac{1}{\gamma_0} \frac{\mathbf{J} \cdot \mathbf{p}_\textit{sig}} {\mathbf{J} \cdot \mathbf{J}}.
\end{equation}
	We can also define a dimensionless ``fidelity''
\begin{equation}\label{gg}
	\mathcal{F} = 1 -\left( \frac{\mathcal{E}^2}{\mathbf{p}_\textit{sig} \cdot \mathbf{p}_\textit{sig}} \right)^{1/2}
\end{equation}
which satisfies $0 \leq \mathcal{F} \leq 1$.
	If the recalled signals scaled by the optimized $\mathcal{G}$ perfectly match the input signals then the fidelity is $\mathcal{F} =1$, while conversely if the recalled signals are completely uncorrelated with the input signals $\mathcal{F} \ll 1$.

\subsection*{Dependence of the gain and fidelity on the friction coefficient and  model}
\textbf{Friction coefficient.}
	Apart from Fig.~5(b) all of the data shown in the main text were taken using the simplest (Cundall-Strack) friction model H.
	When the grain-grain friction coefficient $\mu$ was varied as shown in Fig.~5(a), the resulting gain and fidelity were as follows.

\begin{center}
\begin{tabular}{c | c | c}
$\mu$ & $\mathcal{G}$ & $\mathcal{F}$\\
\hline
\rule{0pt}{10pt}
1 & $1.77 \times 10^5$\ Pa & 0.910 \rule{0pt}{2.5ex}\\  
0.5 & $1.85 \times 10^5$\ Pa & 0.676 \rule{0pt}{2.5ex}\\
0.2 & $2.18 \times 10^5$\ Pa & 0.229 \rule{0pt}{2.5ex}\\
0 & $-1.18 \times 10^5$\ Pa & 0.050 \rule{0pt}{2.5ex}
\end{tabular}
\end{center}	
	Each $\mathcal{G}$ value was determined using all $N=6$ input signals as described above, even though Fig.~5(a) only shows the results for input signal $J_5$.
	
	Examining this table it can be seen that the fidelity $\mathcal{F}$ decreases monotonically with decreasing $\mu$, becoming close to zero when grain-grain friction is turned off ($\mu=0$).
	Conversely the best-fit gain $\mathcal{G}$ does not have a clear trend with varying $\mu$, but rather has a similar magnitude for all $\mu$ values.
	This indicates that as the friction coefficient is reduced, the magnitude of responses to the input signals remains roughly constant while the response waveforms become systematically less correlated with the input waveforms.
	In the limit $\mu \rightarrow 0$ the responses are effectively noise, which happens to be negatively correlated with the inputs for the particular random realization of the grain pack used in this case (indicated by $\mathcal{G}<0$).
	
~\\ \textbf{Friction model.}	
For the data shown in Fig.~5(b) the friction model was varied, keeping all other parameters unchanged including $\mu=1$.
	Hence the top line of this table is the same as the top line of the table above.
\begin{center}
\begin{tabular}{c | c | c}
Model & $\mathcal{G}$ & $\mathcal{F}$\\
\hline
H & $1.77 \times 10^5$\ Pa & 0.910 \rule{0pt}{2.5ex}\\
M1 & $4.53 \times 10^5$\ Pa & 0.460 \rule{0pt}{2.5ex}\\
M2 & $5.46 \times 10^5$\ Pa & 0.891 \rule{0pt}{2.5ex}
\end{tabular}
\end{center}
	As noted in the main text, the memory effect is strong ($\mathcal{F} \approx 0.9$) for models H and M2 while it is considerably weaker for model M1 with $\mathcal{F} = 0.46$.
	As shown in Fig.~5(b) this $\mathcal{F}$ value corresponds to a markedly smaller response to the most complex input signal used, $J_6$.
	A possible physical interpretation is that model M1 smears the response to the input at each $u$ value across a range of $u$ values in the recalled signal.
	
	It can also be seen that the gain $\mathcal{G}$ is about three times larger for model M2 than for model H.
	This is not surprising as the friction force laws (Eqs.~2 and 3 in the main text) are different, but the magnitude and direction of the difference in $\mathcal{G}$ values are not explained.